\documentclass[fleqn,twoside]{article}

\usepackage{espcrc2}
\usepackage{calc}
\usepackage{cite}
\usepackage[english]{babel}
\usepackage[small]{caption2}
\usepackage{graphicx}

\newcommand{\M}[3]{{\mathcal{B}}^{(#1)}_{#2#3}}
\renewcommand{\*}{\,}
\newcommand{\ice}[1]{\relax}
\title{
\vspace*{-3cm}
\centerline{\normalsize\hfill  SSFB/CPP-05-96}
\centerline{\normalsize\hfill  TTP05-24     }
\vspace*{1cm}
Recent progress in computing four-loop massive correlators
\thanks{Talk given at the 12th International Conference on Quantum
Chromodynamics, Montpellier, 4-8th July 2005}}

\author{K.~G.~Chetyrkin\address[KUNI]{
	 Institut f\"ur Theoretische Teilchenphysik,
	 Universit\"at Karlsruhe, 
	 D-76128 Karlsruhe, 
	 Germany
	 }%
         \thanks{On leave from Institute for Nuclear Research of the 
     	         Russian Academy of Sciences, Moscow, 117312, Russia.},
	 J.~H.~K\"uhn\addressmark[KUNI]
	 and C.~Sturm\addressmark[KUNI]%
  \address[TUNI]{Dipartimento di Fisica Teorica, 
  Universit{\`a} di Torino, I-10125 Torino, Italy \& INFN, Sezione di Torino}
}

\begin{document}
\begin{abstract}
We report about recent progress in computing four-loop massive correlators.
The expansion of these correlators in the external momentum leads to
vacuum integrals. The calculation of these vacuum integrals can be used
to determine  Taylor expansion coefficients of the vacuum
polarization function and decoupling functions in perturbative Quantum
chromodynamics.  New results at four-loop order for the lowest Taylor
expansion coefficient of the vacuum polarization function and for the
decoupling relation \ice{function} are presented.
\vspace{1pc}
\end{abstract}
\maketitle
\section{Introduction}
\label{Intro}
Two-point correlators have been studied in great detail in the framework
of perturbative quantum field theory. Due to simple kinematics (only one
external momentum) even multi-loop calculations can be performed. The
results for all physically interesting diagonal and non-diagonal
correlators and including {\em full}  quark   mass dependence 
 are available up to $\mathcal{O}\left(\alpha_s^2\right)$
\cite{Chetyrkin:1996cf,Chetyrkin:1998ix,Chetyrkin:1997mb}.\\
At four-loop order the two-point correlators can be considered in two
limits. In the high energy limit massless propagators need to be
calculated  and in the low energy limit vacuum diagrams
(tadpole integrals without dependence on the external momentum)
arise. The evaluation of these massive tadpoles in three-loop approximation
has been pioneered in ref. \cite{Broadhurst:1992fi} and automated in
ref. \cite{Steinhauser:2000ry}.\\
Similar to the three-loop case, the analytical evaluation of four-loop tadpole
integrals is based on the traditional  Integration-By-Parts (IBP) method. 
In contrast to the three-loop case the manual
construction of algorithms to reduce arbitrary diagrams to a small set
of master integrals is replaced by Laporta's algorithm
\cite{Laporta:1996mq,Laporta:2001dd}. 
In this context the IBP identities are generated
with numerical values for the powers of the propagators and the
irreducible scalar products. In  the next step, the resulting system of linear 
equations is then solved in the next step by expressing systematically complicated
integrals in terms of simpler ones. The resulting  solutions are then 
substituted into all the other equations.\\
This reduction has been implemented in an automated {\tt{FORM3}}
\cite{Vermaseren:2000nd,Vermaseren:2002rp} based program in which
partially ideas described in
ref. \cite{Laporta:2001dd,Mastrolia:2000va,Schroder:2002re} have been
implemented. The rational functions in the space-time dimension $d$, which
arise in this procedure,  are simplified with the program {\tt{FERMAT}}
\cite{Lewis}. The automated exploitation of {\em all} symmetries of the
diagrams by reshuffling the powers of the propagators of a given
topology in a unique way strongly reduces  the number of
equations which need to be solved.

In general, the tadpole diagrams encountered during our calculation
contain both massive and massless lines. In contrast, the
computation of the four-loop $\beta$-functions can be reduced to the
evaluation  of four-loop tadpoles  composed of { completely
massive} propagators.  These special cases have been
considered  in \cite{vanRitbergen:1997va,Schroder:2002re,Czakon:2004bu}.

The outline of this paper is as follows. In section \ref{VacPol} we
discuss the calculation of the lowest expansion coefficient of the
vacuum polarization function and present the results at four-loop order
using  methods as described  above. In section \ref{Decoupling} we
present new results for the decoupling relation at four-loop order in
perturbative QCD. Our conclusions are presented in  section
\ref{Conclusion}.

\section{Vacuum polarization function}
\label{VacPol}
The vacuum polarization tensor $\Pi^{\mu\nu}(q)$ is defined as
\begin{equation}
  \Pi^{\mu\nu}(q)=i\*\int dx\,e^{iqx}\langle 0|Tj^\mu(x) j^\nu(0)|0
  \rangle\,, 
\label{correl}
\end{equation}
where $q^{\mu}$ is the external momentum and $j^{\mu}$ is the
electromagnetic current of a heavy quark with mass $m_h$. The tensor
$\Pi^{\mu\nu}(q)$ can be expressed by a scalar function, the vacuum
polarization function $\Pi(q^2)$ through
\begin{equation}
  \label{vacpol}
  \Pi^{\mu\nu}(q)=\left(q^{\mu}\*q^{\nu}-q^2\*g^{\mu\nu}\right)\*\Pi(q^2)+q^{\mu}\*q^{\nu}\*\Pi_L(q^2).
\end{equation}
The longitudinal part $\Pi_L(q^2)$ vanishes  due to the Ward
identity. The polarization function $\Pi(q^2)$ is
related to the experimentally measurable $R$-ratio $R(s)$ through the
dispersion relation:
\begin{equation}
\label{DisRel}
\Pi(q^2)=
\Pi(q^2=0)+{q^2 \over 12\*\pi^2}\int\!ds {R(s)\over s\*(s-q^2)}\,. 
\end{equation}
Performing the $n$-th derivative of eq.~(\ref{DisRel}) with respect to
$q^2$ at $q^2=0$ one obtains the  moments
$\mathcal{M}_n^{\mbox{\footnotesize{exp}}}$,
which can be determined  experimentally: 
\[
\mathcal{M}_n^{\mbox{\footnotesize{exp}}}= \int\!ds{R(s)\over s^{n+1}}=
{12\*\pi^2\over n!}\!\!  \left.\!\left({d \over
dq^2}\right)^n\!\!\Pi(q^2)\right|_{q^2=0}\,.
\]
The derivatives of the polarization function on the rhs are related to 
the Taylor expansion coefficients $\overline{C}_n$:
\begin{equation}
\overline{\Pi}(q^2) = {3\*Q_q^2\over16\*\pi^2}\*\sum_{n\ge0}
                         \overline{C}_n\*z^n
\label{PiExp}
{},
\end{equation}
($z=q^2/(4\*\overline{m}_h^2)$) which can be calculated in  perturbative QCD.
The first and higher derivatives are
important for a precise determination of the charm- and bottom-quark
mass (see e.g. \cite{Kuhn:2001dm}).  But also 
the lowest expansion coefficient $\overline{C}_0$ has
an interesting physical meaning: it relates the coupling of 
electromagnetic interaction in
different renormalization schemes.  In the case of 
QED-on-shell-renormalization
the residue of the photon propagator is one and the electrical charge
$e$ coincides with the classical limit.  If one performs renormalization
in the $\overline{\mbox{MS}}$-scheme one obtains a relation between the
coupling constant of the electromagnetic interaction
$\alpha_{\mbox{\footnotesize{em}}}=e^2/(4\*\pi)$ in QED-on-shell-renormalization
and the coupling constant
$\overline{\alpha}_{\mbox{\footnotesize{em}}}=\bar{e}^2/(4\*\pi)$ in
the $\overline{\mbox{MS}}$-scheme:
\begin{equation}
\label{MSOnshell}
\alpha_{\mbox{\footnotesize{em}}}=
{\overline{\alpha}_{\mbox{\footnotesize{em}}}\over
        1+\overline{e}^2\*\overline{\Pi}(q^2=0)}\,.
\end{equation}
For massive quarks, interacting through gluons, 
$\overline{\Pi}(q^2=0)$
has been evaluated in  ref.~\cite{Chetyrkin:1996cf}.
At three-loop order in perturbative QCD this relation has already been
determined in ref.~\cite{Chetyrkin:1996cf}. For the QED case the corresponding
result  was  calculated  in ref.~\cite{Broadhurst:1992fi}.

\begin{figure}[!ht]
\begin{minipage}[b]{1.5cm}
  \begin{center}
    \includegraphics[height=1.5cm,bb=126 332 460 665]{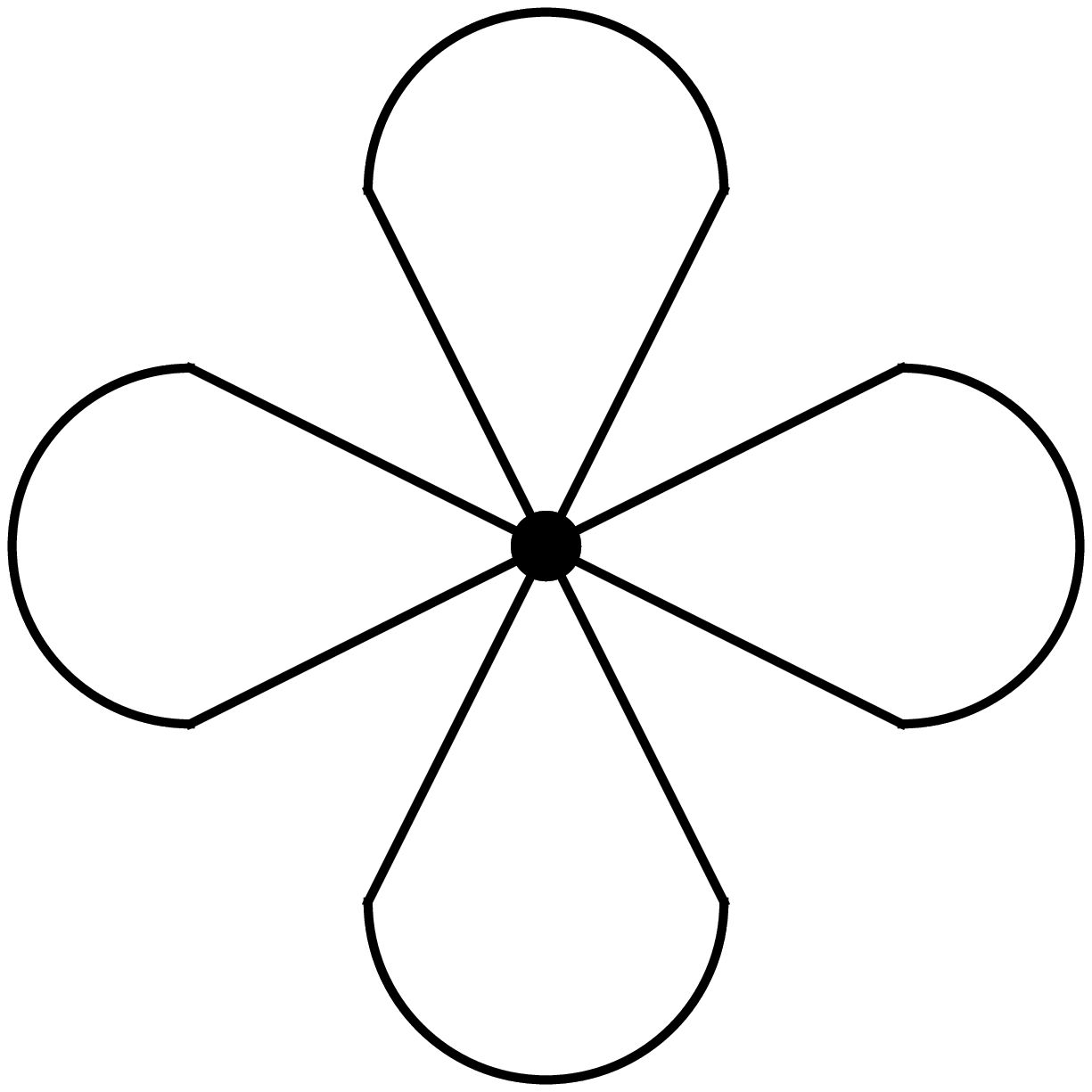}
    $\M{4}{4}{1}$
  \end{center}
\end{minipage}
%
%
\begin{minipage}[b]{1.5cm}
  \begin{center}
    \includegraphics[height=1.5cm,bb=170 320 415 666]{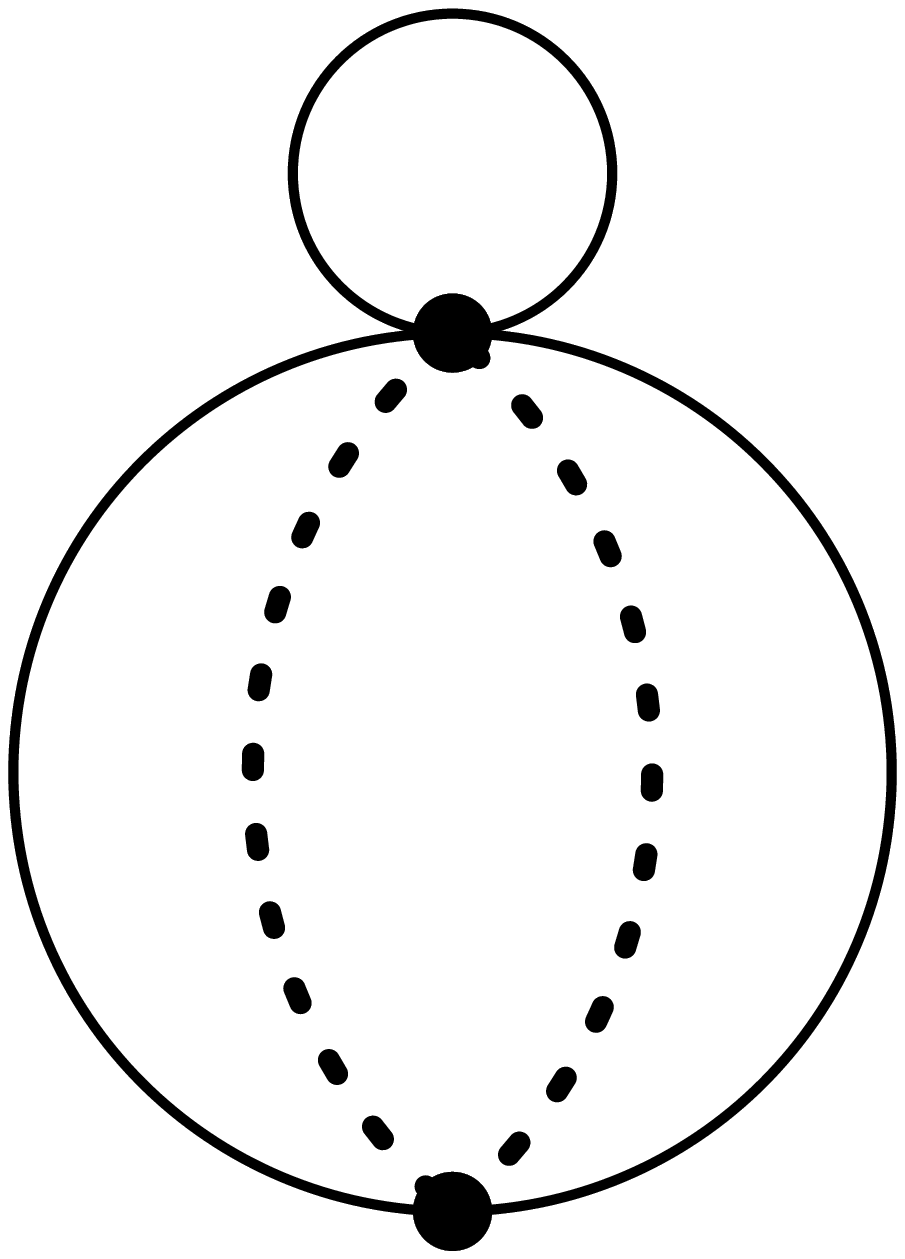}
    $\M{4}{5}{1}$
  \end{center}
\end{minipage}
%
%
\begin{minipage}[b]{1.5cm}
  \begin{center}
    \includegraphics[height=1.5cm,bb=170 320 415 666]{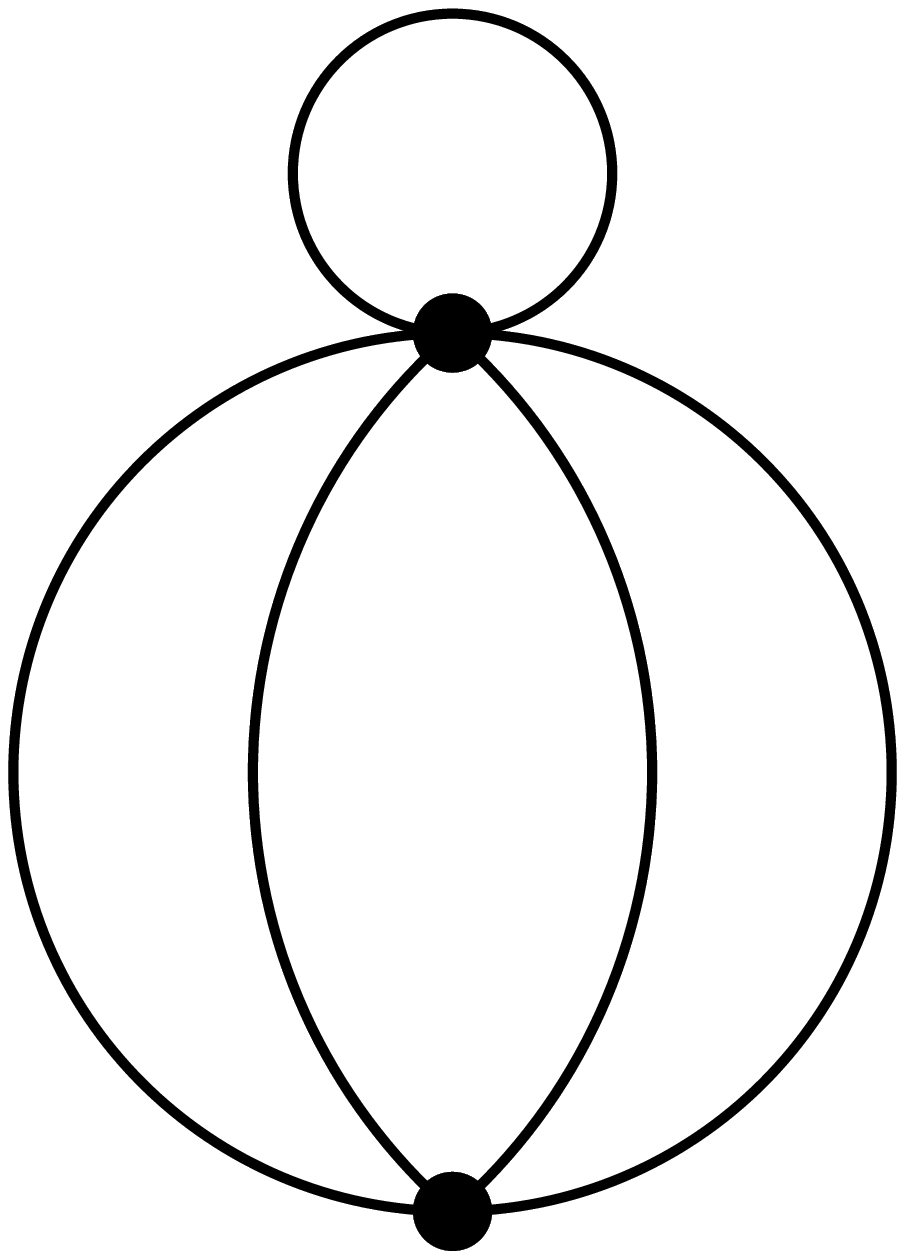}
    $\M{4}{5}{2}$
  \end{center}
\end{minipage}
%
%
\begin{minipage}[b]{1.5cm}
  \begin{center}
    \includegraphics[height=1.5cm,bb=126 320 460 678]{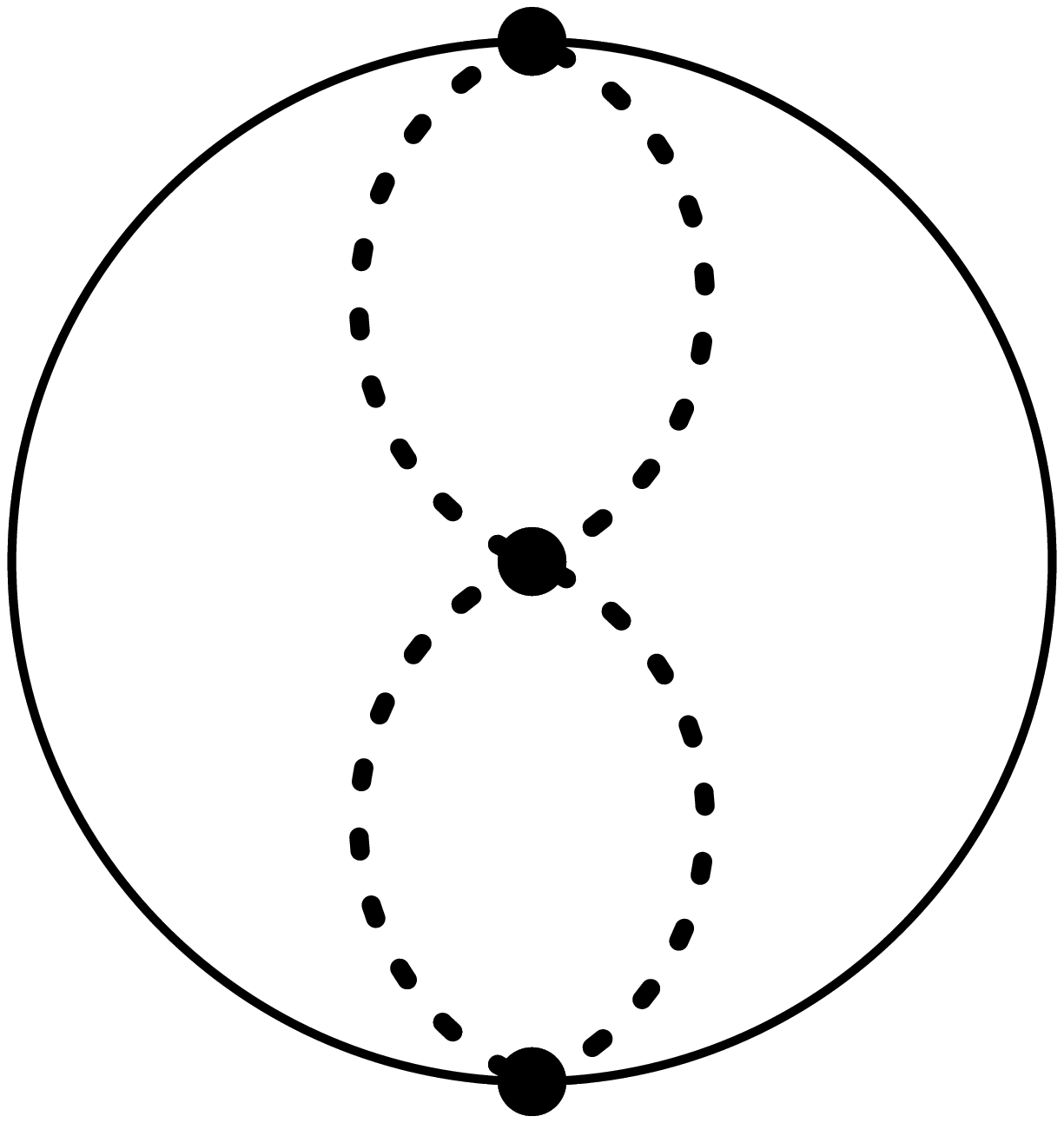}
    $\M{4}{6}{1}$
  \end{center}
\end{minipage}
%
%
\begin{minipage}[b]{1.5cm}
  \begin{center}
    \includegraphics[height=1.5cm,bb=126 320 460 678]{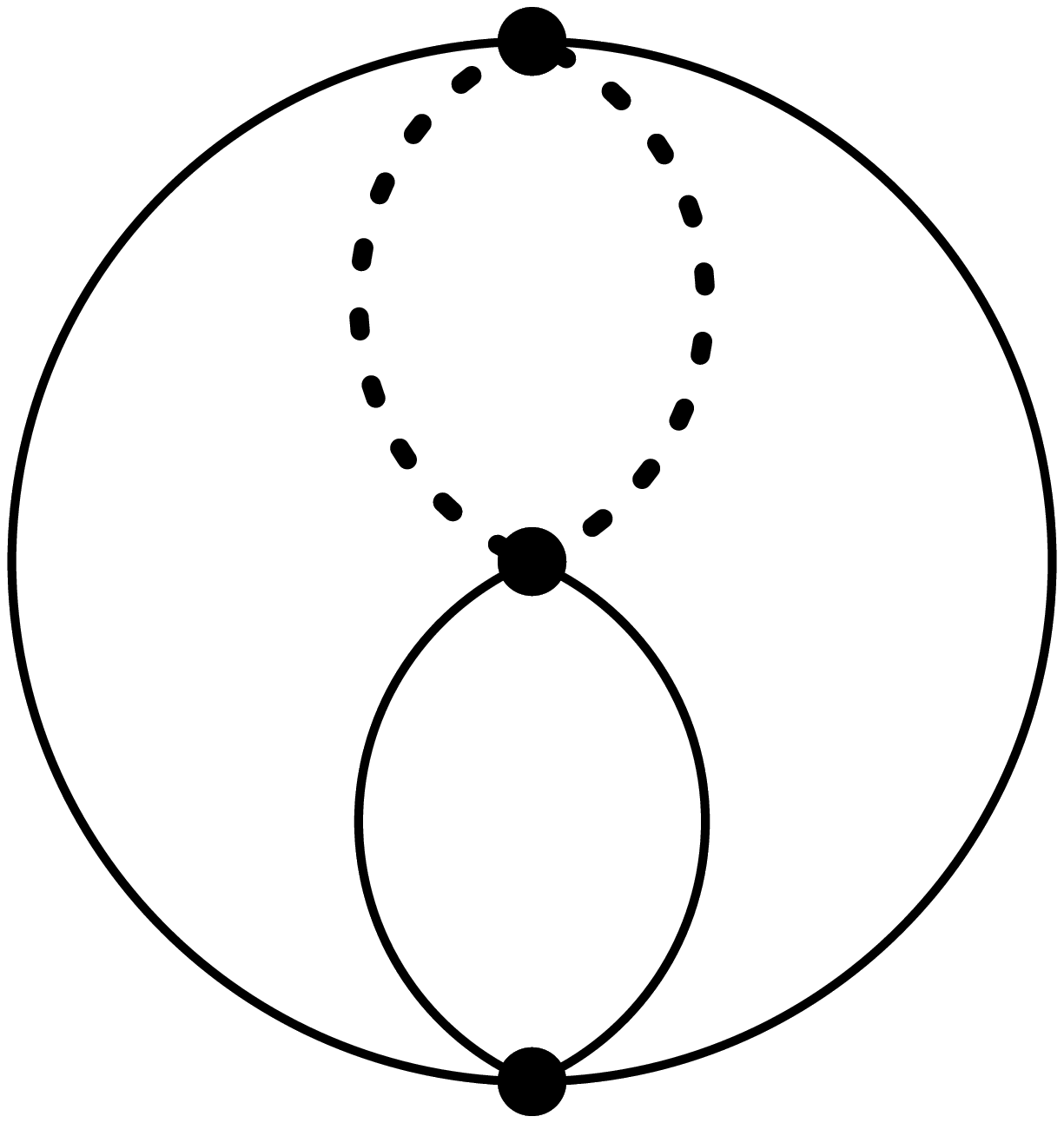}
    $\M{4}{6}{2}$
  \end{center}
\end{minipage}
%
%
\begin{minipage}[b]{1.5cm}
  \begin{center}
    \includegraphics[height=1.5cm,bb=126 320 460 678]{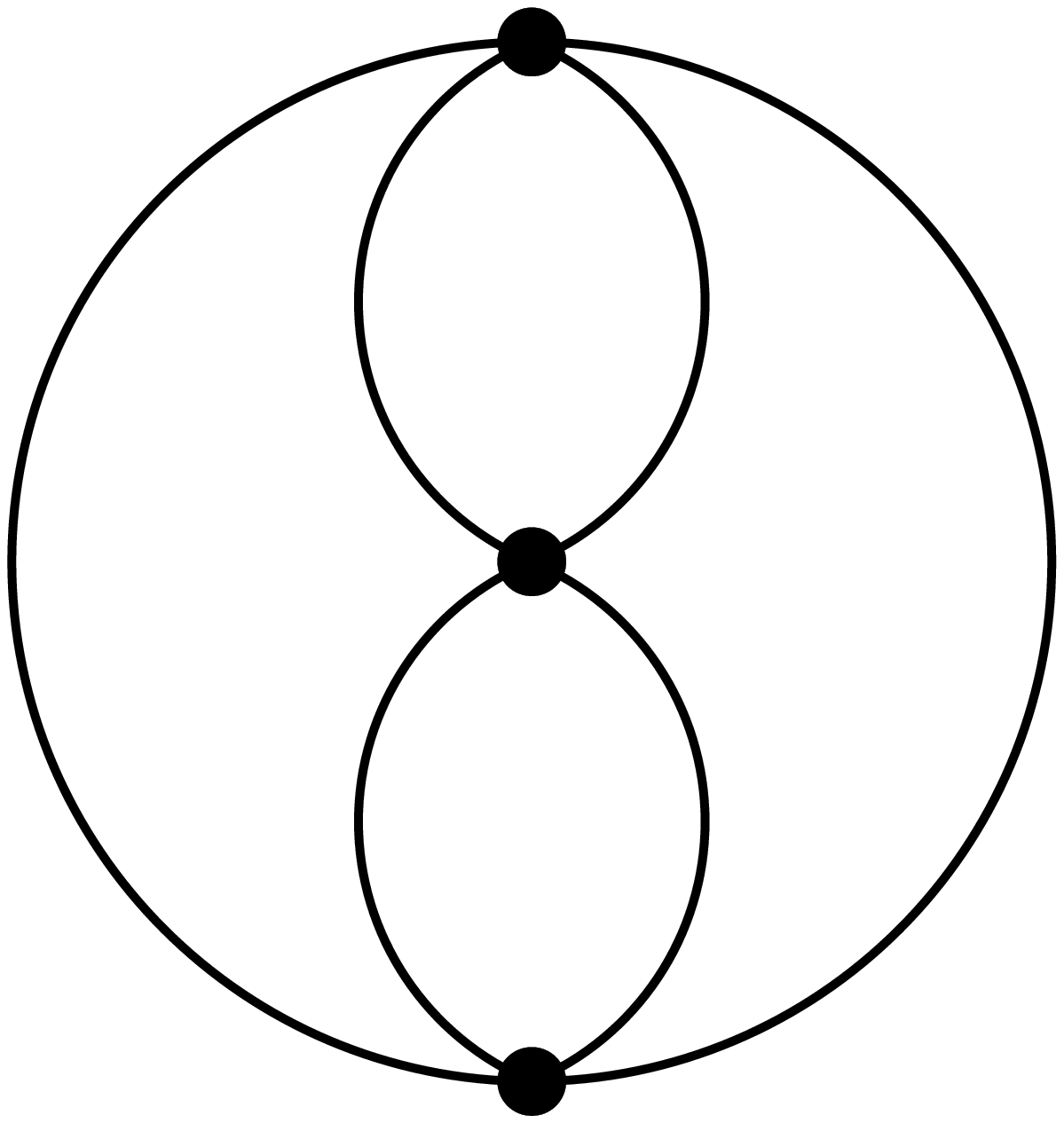}
    $\M{4}{6}{3}$
  \end{center}
\end{minipage}
%
%
\begin{minipage}[b]{1.5cm}
  \begin{center}
    \includegraphics[height=1.5cm,bb=126 320 460 678]{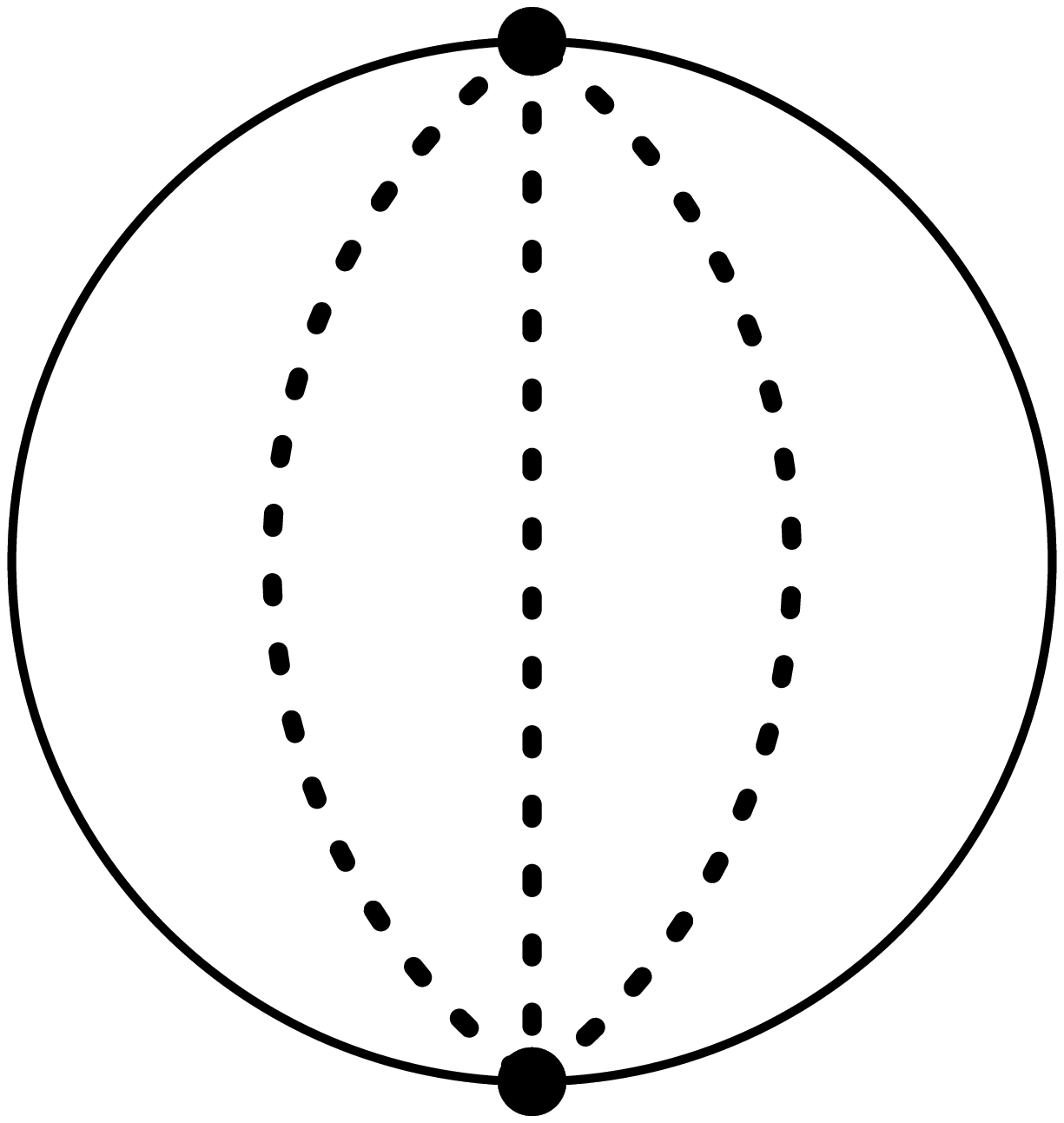}
    $\M{4}{5}{3}$
  \end{center}
\end{minipage}
%
%
\begin{minipage}[b]{1.5cm}
  \begin{center}
    \includegraphics[height=1.5cm,bb=126 320 460 678]{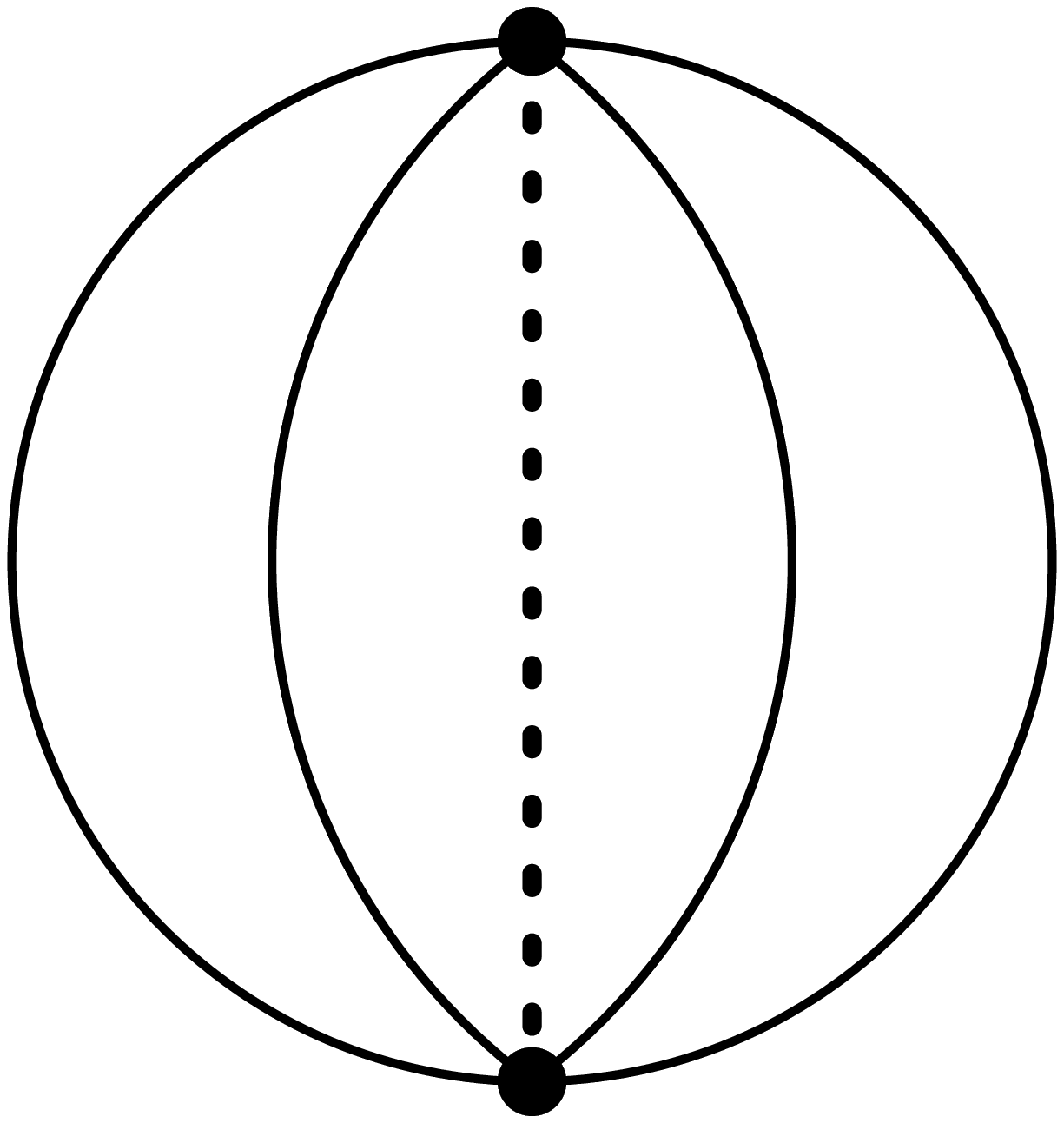}
    $\M{4}{5}{4}$
  \end{center}
\end{minipage}
%
%
\begin{minipage}[b]{1.5cm}
  \begin{center}
    \includegraphics[height=1.5cm,bb=126 332 460 678]{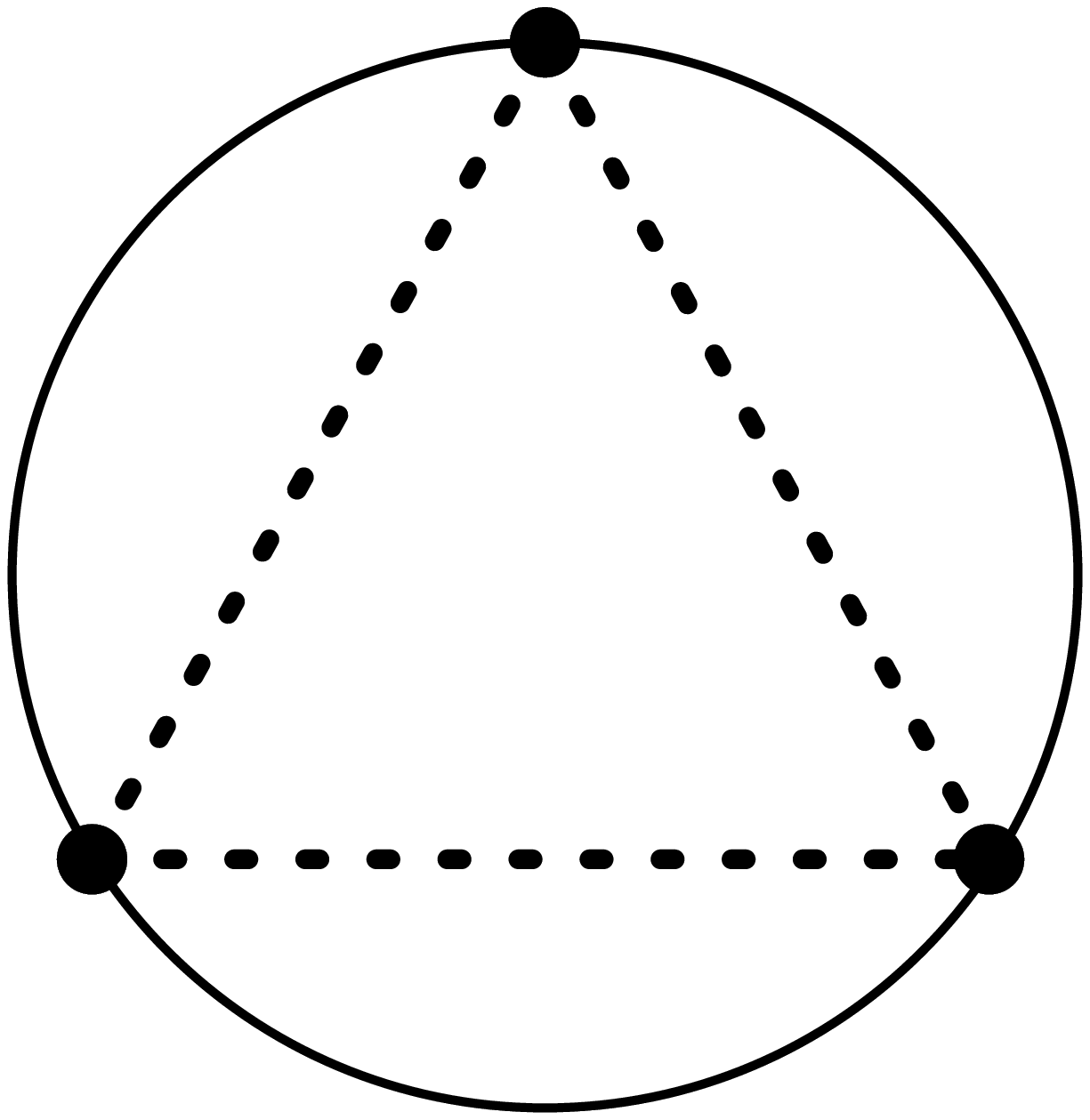}
    $\M{4}{6}{4}$
  \end{center}
\end{minipage}
%
%
\begin{minipage}[b]{1.5cm}
  \begin{center}
    \includegraphics[height=1.5cm,bb=126 332 460 678]{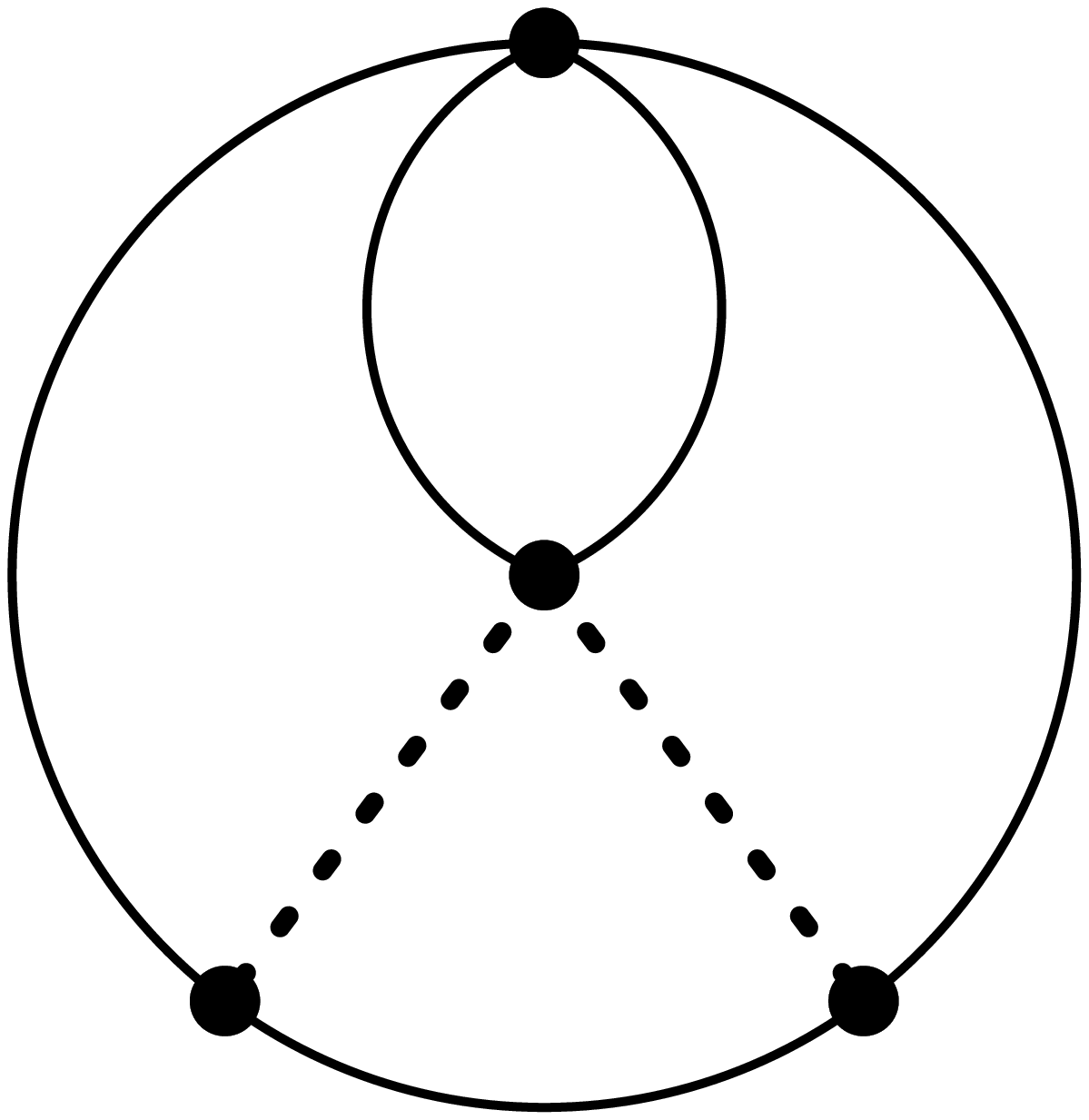}
    $\M{4}{7}{1}$
  \end{center}
\end{minipage}
%
%
\begin{minipage}[b]{1.5cm}
  \begin{center}
    \includegraphics[height=1.5cm,bb=126 332 460 666]{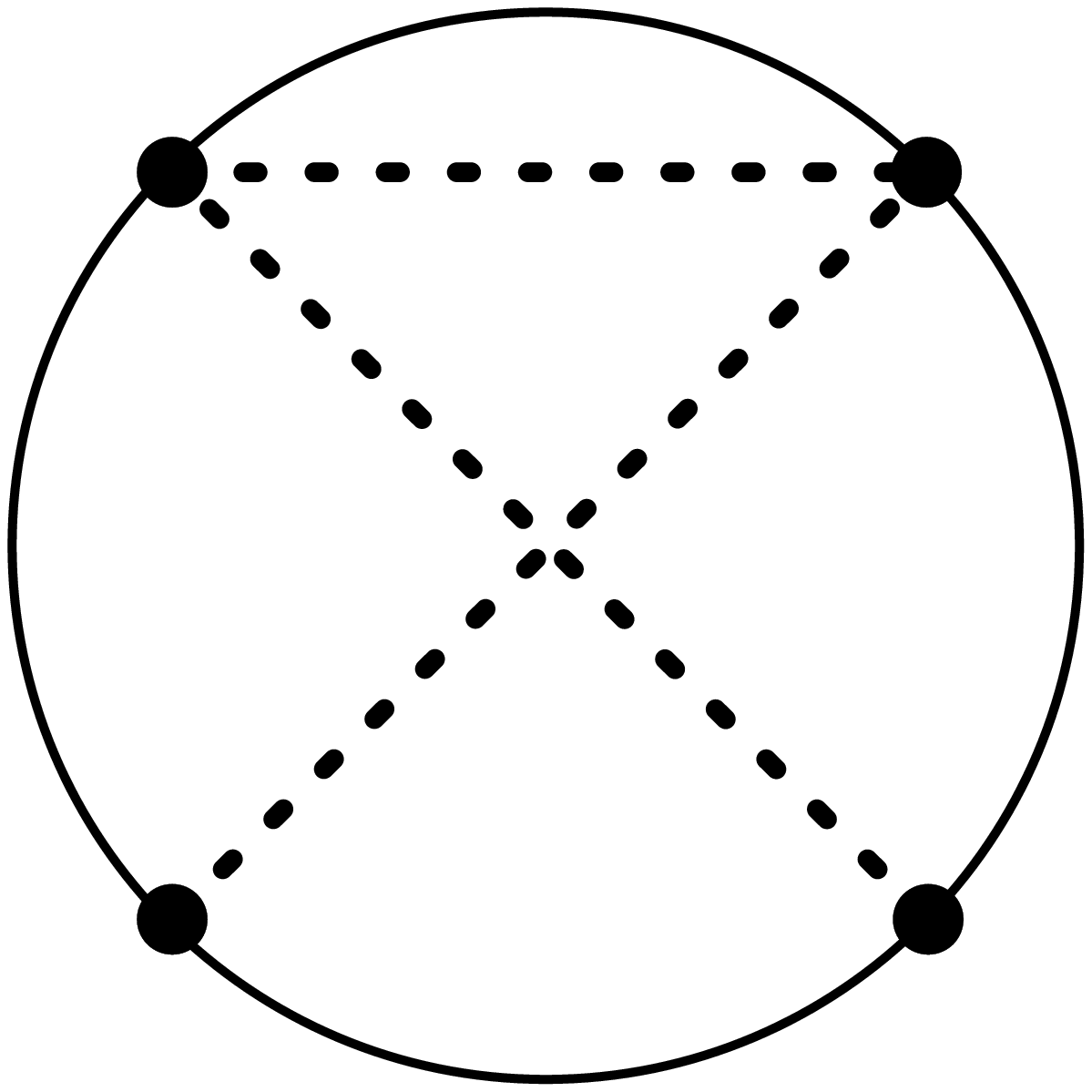}
    $\M{4}{7}{2}$
  \end{center}
\end{minipage}
%
%
\begin{minipage}[b]{1.5cm}
  \begin{center}
    \includegraphics[height=1.5cm,bb=126 332 460 666]{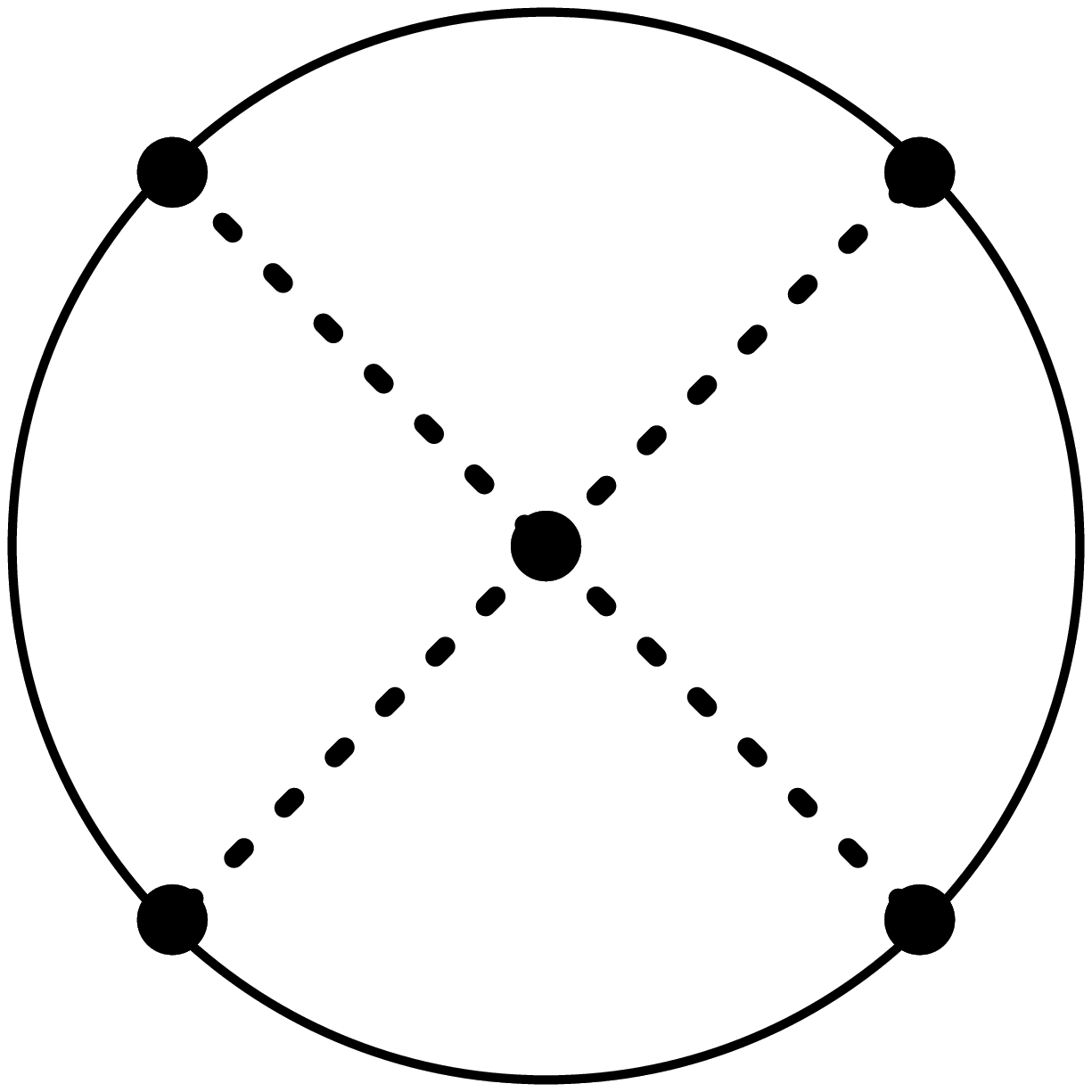}
    $\M{4}{8}{1}$
  \end{center}
\end{minipage}
%
%
\begin{minipage}[b]{1.5cm}
  \begin{center}
    \includegraphics[height=1.5cm,bb=126 320 460 678]{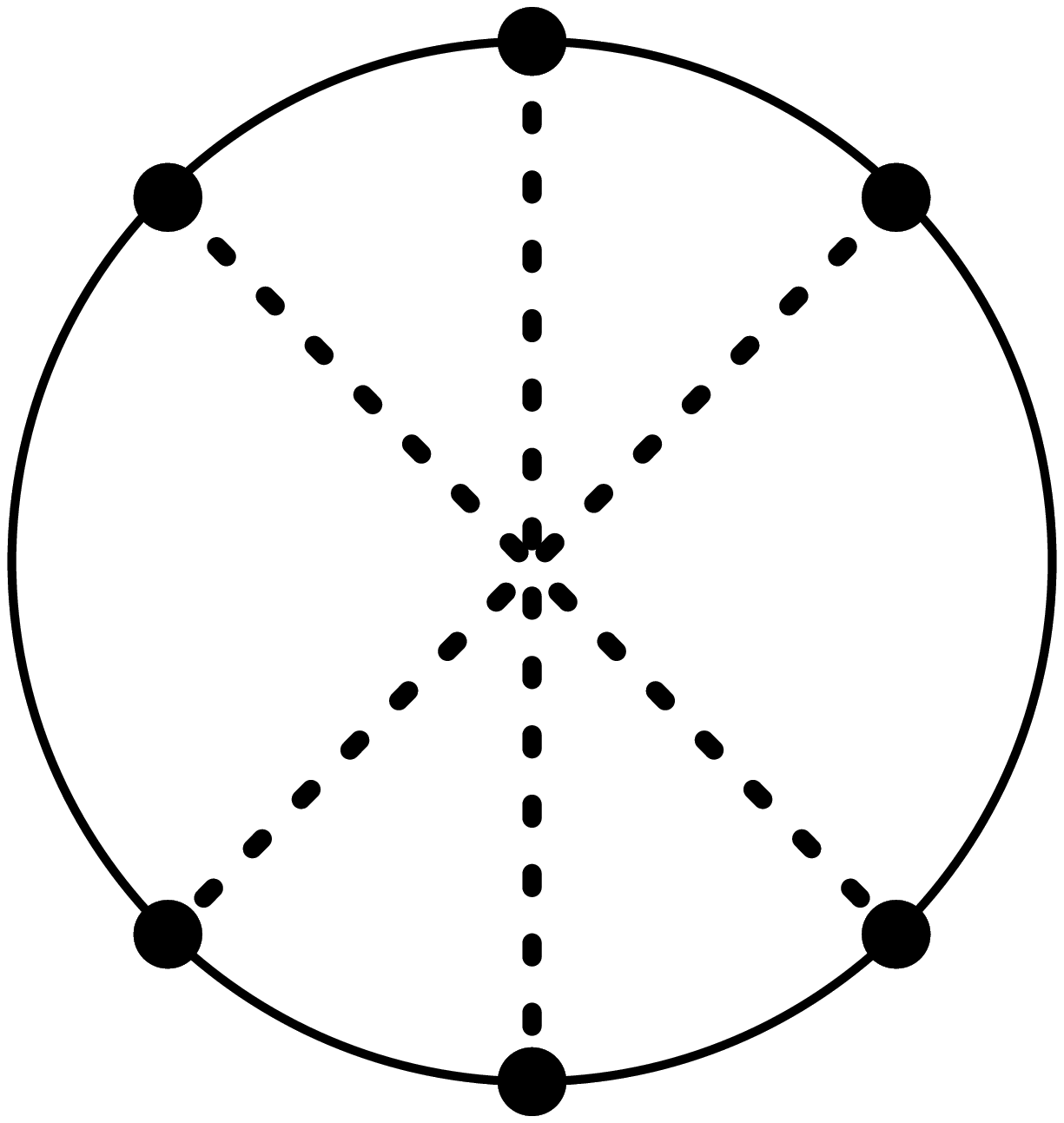}
    $\M{4}{9}{1}$
  \end{center}
\end{minipage}
\caption{Master integrals. The solid lines denote massive lines, whereas
  the dashed lines denote massless lines.\label{MasterTopologies}} 
\end{figure}

The first  Taylor coefficient $\overline{C}_0$ has been calculated using the methods
described in section~\ref{Intro}. All tadpole diagrams were 
expressed through the  set of 13 master integrals shown in figure
\ref{MasterTopologies}.
These master integrals have been calculated in
refs.~\cite{Chetyrkin:2004fq,Schroder:2005va,Chetyrkin:2005}. Inserting 
the master integrals into the lowest Taylor coefficient of the
polarization function and performing the renormalization in
the $\overline{\mbox{MS}}$-scheme one obtains the following
result
\ice{\footnote{The result is also known analytically for the four-loop
contributions being proportional to $n_f$ and will be presented
elsewhere.}}
for the expansion coefficient
$\overline{C}_0(\mu=\overline{m}_h)$:
\begin{eqnarray}
\label{C0Numerik}
\overline{C}_0 
\!\!\!\!&=&\!\!\!\!\left({\alpha_s\over\pi}\right) \* 1.4444
+
\left({\alpha_s\over\pi}\right)^2
    \left( 0.3714\*n_l + 1.725 \right)
\nonumber
\\
\hspace{-4mm}
\!\!\!\!&+&\!\!\!\!\left({\alpha_s\over\pi}\right)^3\*
  ( 0.0257\*n_l^2 - 1.186\*n_l -1.955 )
{},
\end{eqnarray}
where $\alpha_s$ is the strong coupling constant  in the 
$\overline{\mbox{MS}}$-scheme and the symbol $n_l$ denotes the number of
light quarks  considered as massless. 
The part of the four-loop contribution 
proportional to the number $n_f^2$ of active quarks has 
been calculated  previously up to the first physical moment 
$\overline{C}_1$  in
ref.~\cite{Chetyrkin:2004fq}.  The terms   proportional to
$\alpha_s^j\*n_l^{j-1}$ are even known to all orders $j$ 
\cite{Grozin:2004ez}.\\
In ref. \cite{Faisst:2005diss,Faisst:2004kz} the terms being
proportional to $n_f$ have been calculated through  an
independent numerical method, which is based on the Pad{\'e}
approximation
\cite{Fleischer:1994ef,Fleischer:1994dc,Broadhurst:1994qj}. These
results are in complete agreement with those of eq.~(\ref{C0Numerik}).
\section{Decoupling relation}
\label{Decoupling}
The masses of the known quark species differ vastly in their
magnitude. Often the  mass of a
heavy quark $h$ is much larger than the characteristic momentum scale
$\sqrt{s}$ of the reaction under consideration. In such a
two interrelated problems appear  when using an  MS-like
renormalization scheme.\\
First,  two large but  quite different mass scales,
$\sqrt{s}$ and $m_h$, lead to  two different types of potentially
dangerously large logarithms of $\sqrt{s}/\mu$ and $m_h/\mu$ and  
the standard trick of a proper choice of
the renormalization scale $\mu$ is no longer applicable;
 \\
Second, according to the Appelquist-Carazzone theorem
\cite{Appelquist:1974tg} the effects due to heavy particles 
eventually should in general  
`decouple' from  low-energy physics.
However,
a peculiarity of mass-independent renormalization schemes is that the
decoupling theorem does not hold in its naive form for theories
renormalized in such schemes. The  effective QCD action that appears will not be 
not  canonically normalized. Large mass logarithms appear, when 
one calculates a physical observable. \\
Fortunately, both problems are controlled once  the
expansion parameters are properly choosen  and  renormalization group improvement
is performed
\cite{Weinberg:1980wa,Wetzel:1981qg,Bernreuther:1981sg,Bernreuther:1983zp}.
To be specific,  consider QCD with $n_l=n_f-1$ light quarks
and one heavy quark $h$ with mass $m_h$. The effective coupling constant
$\alpha'_s$ is then expressed in terms of the one of the full theory through:
\begin{equation}
\alpha'_s(\mu)=\alpha_s(\mu)\;\zeta_g^2(\alpha_s(\mu),x)\,,\;\; 
x=\log(\mu^2/\overline{m}_h^2)\,,
\end{equation}
where $\zeta_g$  is  the decoupling function and $\overline{m}_h(\mu)$ is 
the $\overline{\mbox{MS}}$ running mass of the heavy quark. The decoupling function is known
up to three-loop order
\cite{Chetyrkin:1997un,Chetyrkin:1997sg}. 
Its calculation  can be reduced to the solution of vacuum
integrals \cite{Chetyrkin:1997un}. Thus the  methods  described in
section~\ref{Intro} can be applied  to calculate the decoupling
function at four-loop order. The same
master integrals as those shown in figure~\ref{MasterTopologies} appear.  
For a renormalization scale $\mu=\overline{m}_h(\mu)$ one obtains the
following numerical result\footnote{After the completion of our calculation we have been
informed that an  independent calculation of 
the matching function $\zeta_g^2$ at the four-loop level has been
recently finished  by Y.~Schr\"oder and  M.~Steinhauser.
The same is true for $\overline{C}_0$ of eq.~(\ref{C0Numerik}).
Their results are in full agreement with ours.
Both  calculations  have been made reecntly  available 
in \cite{Chetyrkin:2005ia,Schroder:2005hy}.
}:
\begin{eqnarray}
\label{CNumerik}
\nonumber
\zeta_g^2&&\!\!\!\!\!\!\!\!\!\!=
1 + \left({\alpha_s\over\pi}\right)^2\*0.1528
\\
&&\!\!\!\!\!\!\!\!\!\!\!
+\left({\alpha_s\over\pi}\right)^3\*\left(0.9721 - 0.0847\*n_l\right)
\\
\nonumber
&&\!\!\!\!\!\!\!\!\!\!\!+ \left({\alpha_s\over\pi}\right)^4\*
  \left(5.1703 - 1.001\*n_l-0.0220\*n_l^2\right)
\,.
\end{eqnarray}
The knowledge of this decoupling function is of phenomenological
importance, because it allows the determination of $\alpha_s(M_Z)$ at the
$Z$-Boson scale through evolution of the measured value of
$\alpha_s(m_{\tau})$ at the $\tau$-lepton scale.  A careful analysis of
the effects of four-loop running and three-loop matching on the  extraction of 
$\alpha_s^{(5)}(M_Z)$ defined for  5 active quark flavors from
$\alpha_s^{(3)}(m_{\tau})$ defined for  3 active quark flavors has been recently 
performed \cite{Davier:2005xq}. We have checked that the  inclusion  of 
the newly computed  four-loop matching
condition leads to a further reduction of the theoretical error from the
evolution.
\section{Conclusion}
\label{Conclusion}
We have calculated the
lowest Taylor expansion coefficient of the vacuum polarization function
and the decoupling relation at four-loop order in perturbative QCD. The
lowest Taylor expansion coefficient  relates the electromagnetic  
coupling in the on-shell  and in the $\overline{\mbox{MS}}$
renormalization schemes. The decoupling relation is important for the
determination of $\alpha_s^{(5)}(M_Z)$ through evolution from the
measured value of $\alpha_s^{(3)}(m_{\tau})$.\\

\noindent
{\bf Acknowledgments:}\\ 
We would like to thank M. Faisst for discussions about the Pad{\'e}
approach and M. Tentyukov for great help in dealing with {\tt FORM3} and
{\tt FERMAT}. K.~Ch. is grateful to  M.  Steinhauser 
for useful discussions of the decoupling relations.
C.S. would like to thank the Graduierten Kolleg 
"{\it{Hochenergiephysik und Teilchenastrophysik}}" for financial
support. This work was supported by the Sonderforschungsbereich
Transregio 9.

\end{document}